\documentclass[12pt]{article}
\usepackage[utf8]{inputenc}

\usepackage[margin=1in]{geometry}
\usepackage{complexity}
\usepackage{mathtools,amsmath,amssymb}
\usepackage{scrextend}
\usepackage{color,subfig,}
\usepackage{algorithmicx}
\usepackage{algorithm}
\usepackage[noend]{algpseudocode}
\usepackage{cite}

\algnewcommand\algorithmicinput{\textbf{Input:}}
\algnewcommand\Input{\item[\algorithmicinput]}

\algnewcommand\algorithmicoutput{\textbf{Output:}}
\algnewcommand\Output{\item[\algorithmicoutput]}

\algnewcommand\algorithmictask{\textbf{Task:}}
\algnewcommand\Task{\item[\algorithmictask]}

\algnewcommand\algorithmicgoal{\textbf{Goal:}}
\algnewcommand\Goal{\item[\algorithmicgoal]}

\usepackage{footnote}
\newcommand{\astfootnote}[1]{
\let\oldthefootnote=\thefootnote
\setcounter{footnote}{0}
\renewcommand{\thefootnote}{\fnsymbol{footnote}}
\footnote{#1}
\let\thefootnote=\oldthefootnote
}

\usepackage{comment}
\usepackage{todonotes}
\usepackage{xspace}
\usepackage{multirow}
\usepackage{xcolor}
\RequirePackage[colorlinks=true]{hyperref}
\hypersetup{
	linkcolor=[rgb]{0,0,0.5},
	citecolor=[rgb]{0, 0.5, 0},
	urlcolor=[rgb]{0.5, 0, 0}
}
\usepackage{dsfont}

\usepackage{tikz}
\usepackage{tcolorbox}
\usetikzlibrary{positioning}
\usetikzlibrary{shapes,arrows}
\usetikzlibrary{shadows,arrows}

\newtcolorbox[use counter = box]{pabox}[2][]{fonttitle=\bfseries,
title=Box~\thetcbcounter: #2,#1}

\usepackage{amsthm}
\usepackage{thmtools,thm-restate}

\numberwithin{equation}{section}
\declaretheoremstyle[bodyfont=\it,qed=\qedsymbol]{noproofstyle}

\declaretheorem[name=Observation,numbered=no]{observation*}

\declaretheorem[name=Theorem,numbered=no]{theorem*}

\declaretheorem[name=Lemma,numbered=no]{lemma*}

\declaretheorem[name=Corollary,numbered=no]{corollary*}

\declaretheorem[name=Proposition,numbered=no]{proposition*}

\declaretheorem[name=Claim,numbered=no]{claim*}

\declaretheorem[name=Conjecture,numbered=no]{conjecture*}

\declaretheorem[name=Question,numbered=no]{question*}

\declaretheoremstyle[bodyfont=\it]{defstyle}

\declaretheorem[unnumbered,name=Definition,style=defstyle]{definition*}

\declaretheorem[unnumbered,name=Example,style=defstyle]{example*}

\declaretheorem[unnumbered,name=Notation=defstyle]{notation*}

\declaretheorem[unnumbered,name=Construction,style=defstyle]{construction*}

\declaretheoremstyle[]{rmkstyle}

\newcommand{\Real}{\mathbb{R}}

\makeatletter
\def\moverlay{\mathpalette\mov@rlay}
\def\mov@rlay#1#2{\leavevmode\vtop{%
   \baselineskip\z@skip \lineskiplimit-\maxdimen
   \ialign{\hfil$\m@th#1##$\hfil\cr#2\crcr}}}
\newcommand{\charfusion}[3][\mathord]{
    #1{\ifx#1\mathop\vphantom{#2}\fi
        \mathpalette\mov@rlay{#2\cr#3}
      }
    \ifx#1\mathop\expandafter\displaylimits\fi}
\makeatother

\renewcommand{\epsilon}{\varepsilon}

\renewcommand{\qedsymbol}{\hfill{\Large $\square$}}

\renewcommand{\P}{\mathbb{P}}

\counterwithout{equation}{section}

\author{Mien Brabeeba Wang\footnotemark[1]~\footnotemark[2] and Michael M. Halassa\footnotemark[2]}

\title{Thalamocortical contribution to solving  credit assignment in neural systems}

\begin{document}
\maketitle
\footnotetext[1]{Computer Science \& Artificial Intelligence Laboratory, Massachusetts Institute of Technology, Cambridge, MA.}
\footnotetext[2]{Department of Brain and Cognitive Science, Massachusetts Institute of Technology, Cambridge MA.}

\begin{abstract}
Animal brains evolved to optimize behavior in dynamically changing environments, selecting actions that maximize future rewards. A large body of experimental work indicates that such optimization changes the wiring of neural circuits, appropriately mapping environmental input onto behavioral outputs. A major unsolved scientific question is how optimal wiring adjustments, which must target the connections responsible for rewards,  can be accomplished when the relation between sensory inputs, action taken, environmental context with rewards is ambiguous. The computational problem of properly targeting cues, contexts and actions that lead to reward is known as \textit{structural, contextual and temporal credit assignment} respectively. In this review, we survey prior approaches to these three types of problems and advance the notion that the brain's specialized neural architectures provide efficient solutions. Within this framework, the thalamus with its cortical and basal ganglia interactions serve as a systems-level solution to credit assignment. Specifically, we propose that thalamocortical interaction is the locus of meta-learning where the thalamus provides cortical control functions that parametrize the cortical activity association space. By selecting among these control functions, the basal ganglia hierarchically guide thalamocortical plasticity across two timescales to enable meta-learning. The faster timescale establishes contextual associations to enable rapid behavioral flexibility while the slower one enables generalization to new contexts. Incorporating different thalamic control functions under this framework clarifies how thalamocortical-basal ganglia interactions may simultaneously solve the three credit assignment problems. 

\end{abstract}

\paragraph{Introduction} Learning which action to choose in an uncertain environment is a hallmark of intelligence~\cite{Thorndike2017,Miller2001,Niv2009}. When animals explore unfamiliar environments, they tend to reinforce actions that lead to unexpected rewards. A common notion in contemporary neuroscience is that such behavioral reinforcement emerges from changes in synaptic connectivity, where synapses that contribute to the unexpected reward are strengthened~\cite{Hebb2002,Bliss1973, Abbott2000, Dayan2005, Whittington2019}. A prominent model for connecting synaptic to behavioral reinforcement is dopaminergic innervation of basal ganglia (BG), where dopamine (DA) carries the reward prediction error (RPE) signals to guide synaptic learning~\cite{Montague1996, Schultz1997, Bayer2005, Bamford2018}. This circuit motif is thought to implement a basic form of the reinforcement learning algorithm~\cite{Houk1994, Suri1999, Sutton1990,Wickens1994,Morris2006,Roesch2007, Sutton2018}, which has had much success in explaining simple Pavlovian and instrumental conditioning~\cite{Sutton1990,Ikemoto1999, Niv2009, Sutton2018}. However, what allows this circuit to reinforce the appropriate connections in complex natural environments where animals are presented with multiple cues in multiple contexts and make multiple actions before they receive the reward, is unknown. If one naively credits all synapses with the RPE signals, the learning will be highly inefficient since different cues, contexts and actions contribute to the RPE signals differently. To properly credit the cues, context and actions that lead to unexpected reward is a challenging problem, known as the \textit{credit assignment} problem \cite{Minsky1961, Rumelhart1986, Whittington2019, Lillicrap2020}. 

\begin{figure}[h!]
    \centering
    \includegraphics[width = 0.95\textwidth]{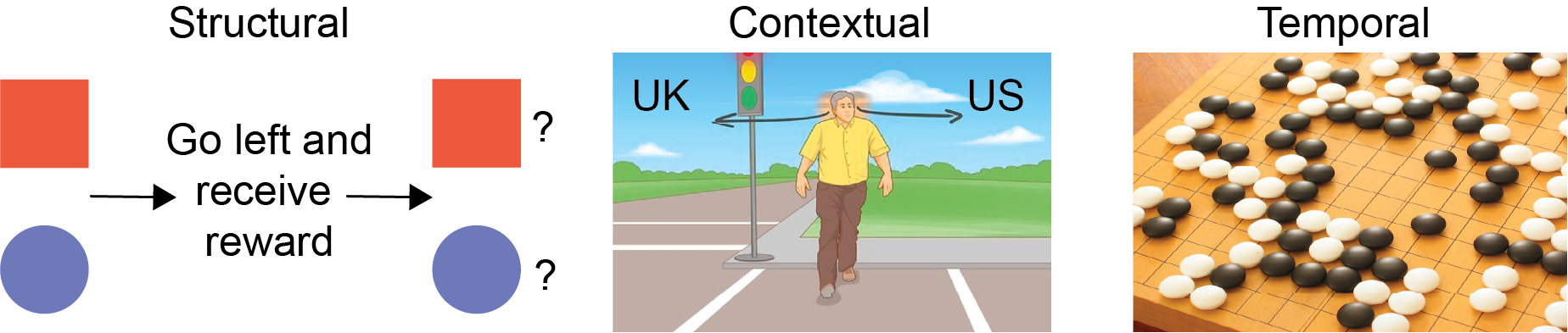}
    \caption{\textbf{Different types of credit assignment} When animals face a decision with multiple cues, contexts and actions before the reward, animals need to assign proper credit to the cues, contexts and actions that lead to behavioral outcomes. In structural credit assignment, animals need to credit the cues that lead to the reward but not the distractor cues. In contextual credit assignment, animals need to credit sensory cues and actions in the right context. In temporal credit assignment, animals need to figure out which past actions lead to the current reward.}
    \label{fig:credit}
\end{figure}

One can roughly categorize the credit assignment into \textit{structural credit assignment}, \textit{contextual credit assignment} and \textit{temporal credit assignment} (\autoref{fig:credit}). In structural credit assignment, animals may make decisions in a multi-cue environment and should be able to credit those cues that contribute to the rewarding outcome. Similarly, if actions are being chosen based on internal decision variables, then the underlying activity states must also be reinforced. In such cases, neurons that are selective to external cues or internal latent variables need to adjust their downstream connectivity based on its contribution of their downstream targets to the RPE. This is a challenging computation to implement because, for upstream neurons, the RPE will be dependent on downstream neurons that are several connections away. For example, a sensory neuron needs to know the action chosen in the motor cortex to selectively credit the sensory synapses that contribute to the action. In contextual credit assignment, animals not only need to appropriately credit the sensory cues and actions that lead to the reward but also need to credit the sensorimotor combination in the right context. For example, when one is in the United States, one learns to first look left before crossing the street, whereas, in the United Kingdom, one learns to look right instead. However, after spending time in the UK, someone from the US should not unlearn the behavior of looking left first when they return home because their brain ought to properly assign the credit to a different context. In the temporal credit assignment problem, animals make decisions in an environment with distant rewards and need to figure out which past sensory cues and actions lead to the current reward. For example, in a game of Go, even though the result of the game is only revealed after hundreds of hands, professional players can recognize which moves in the past are good and reinforce such moves.

In this review, we will first go over common approaches from machine learning to tackle these three credit assignment problems. In doing so, we highlight the challenge in their efficient implementation within biological neural circuits. We also highlight some recent proposals that advance the notion of specialized neural hardware that approximate more general solutions for credit assignment \cite{OReilly1996,Roelfsema2005, Fiete2006, Ketz2013, Lillicrap2016, Schiess2016, Kusmierz2017, Zenke2018, Richards2019, Roelfsema2018, Sacramento2018, Kornfeld2020,Liu2020,oreilly2021}. Along these lines, we propose an efficient systems-level solution involving the thalamus and its interaction with the cortex and BG for these three credit assignment problems.

\paragraph{Common machine learning approaches to credit assignment} One solution to structural credit assignment in machine learning is backpropagation \cite{Rumelhart1986}. Backpropagation recursively computes the vector-valued error signal for synapses based on their contribution to the error signal. There is much empirical success of backpropagation in surpassing human performance in supervised learning such as image recognition \cite{Krizhevsky2012,He2016} and reinforcement learning such as playing the game of Go and Atari \cite{Mnih2015, Silver2016, Silver2017,Schrittwieser2020}. Additionally, comparing artificial networks trained with backpropagation to neural responses from the ventral visual stream of non-human primates shows comparable internal representations  \cite{Cadieu2014,Yamins2014}. Despite its empirical success in superhuman level performance and matching the internal representation of actual brains, backpropagation may not be straightforward to implement in biological neural circuits as we explain below.

In its most basic form, backpropagation requires symmetric connections between neurons (forward and backward connections). Mathematically, we can write down the backpropagation in \autoref{eq: back prop}:
\begin{equation}
\label{eq: back prop}
\delta W_i \propto \frac{\partial E}{\partial W_i} = e_if(a_{i-1})^\top 
\end{equation}
where
\begin{equation*}
e_i = W_{i+1}^\top e_{i+1} \circ f'(a_i),  
\end{equation*}
$E$ is the total error, $e_i$ is the vector error at layer $i$, $W_i$ is the synaptic weight connecting layer $i-1$ to layer $i$ and $f$ is the nonlinearity. Intuitively, this is saying that the change of synaptic weight $W_i$ is computed by a Hebbian learning rule between backpropagation error $e_i$ and activity from last layer $f(a_{i-1})$ while the backpropagation error is computed by backpropagating the error in the next layer through symmetric feedback weights $W_{i+1}^\top$. Importantly, in this algorithm, error signals do not alter the activity of neurons in the preceding layers and instead operate independently from the feedforward activity. However, such arrangement is not observed in the brain; symmetric connections across neurons are not a universal feature of circuit organization, and biological neurons may encode both feedforward inputs and errors through changes in spike output (changes in activity)~\cite{Crick1989,Richards2019}. Therefore, it is hard to imagine how the basic form of backpropagation (symmetry and error/activity separation) is physically implemented in the brain.

Furthermore, while an animal can continually learn to behave across different contexts, artificial neural networks trained by backpropagation struggle to learn and remember different tasks in different contexts: a problem known as catastrophic forgetting \cite{McCloskey1989, French1999,Kumaran2016, Kemker2018, Parisi2019}. Specifically, this problem occurs when the tasks are trained sequentially because the weights optimized for former tasks will be modified to fit the later tasks. One of the common solutions is to interleave the tasks from different contexts to jointly optimize performance across contexts by using an episodic memory system and replay mechanism~\cite{McClelland1995, Kumaran2016}. This approach has received empirical success in artificial neural networks including learning to play many Atari game~\cite{Mnih2015, Schrittwieser2020}. However, since one needs to store past training data in memory to replay during learning, this approach demands a high computational overhead and can be is inefficient as the number of the contexts increases. On the other hand, humans and animals acquire diverse sensorimotor skills in different contexts throughout their life span: a feat that cannot be solely explained by memory replay~\cite{Murray2016, Power2017, Parisi2019,Zenke2017}. Therefore, biological neural circuits are likely to employ other solutions to contextual credit assignment in addition to memory replay. 

Lastly, one of the most common solutions to temporal credit assignment in reinforcement learning is temporal difference (TD) learning \cite{Sutton1990, Sutton2018}. Instead of directly learning from distant rewards, animals use more proximal inputs (which form a current state) to learn the expected future rewards. Specifically, the expected future reward can be express as
\begin{equation}
V(s_t) = \mathbb{E}\left[r_t + r_{t+1} + r_{t+2} + \dotsb \middle| s_t\right] =  \mathbb{E}\left[\sum_{i=0}^\infty r_{t+i} \middle| s_t\right],
\end{equation}
where $s_t$ is the state at time $t$ and $r_t$ is the reward at time $t$. The learning is guided by the temporal prediction error
\begin{equation}
e_t = r_t + V(s_{t+1}) - V(s_t).
\end{equation}
TD learning brings the distant reward to the present by estimating the future rewards given the current state, and thus solve the temporal credit assignment problem by reinforcing the action that leads to positive TD RPE. However, for TD learning to be valid, state transition and rewards need to depend on only the current state and the current action. In natural settings, both rewards and sensory inputs can depend on past sensory information and past actions. For example, when people communicate with each other, their intentions may only be clear after a whole sentence, not the individual words spoken. Thus, the naive implementation of TD learning on sensory inputs can fail on such tasks with sequential dependency. To overcome this shortcoming, a working memory buffer may be required, and in machine learning can be implemented via Long Short-Term Memory networks (LSTMs) \cite{Hochreiter1997, Yu2019, VanHoudt2020}. An LSTM can selectively maintain the past sensory inputs in hidden memory states. However, one needs to use backpropagation to train LSTM which is unlikely to be implemented in the brain as such. The brain probably has an alternative mechanism to selectively maintain the past sensory inputs and actions in conjunction with TD RPE signals from DA to solve temporal credit assignment.

Therefore, to solve the three credit assignment problems in the brain, one needs to seek different solutions. One of the pitfalls of backpropagation is that it is a general algorithm that works on any architecture. However, actual brains are collections of specialized hardware put together in a specialized way. It can be conceived that through clever coordination between different cell types and different circuits, the brains can solve the credit assignment problem by leveraging its specialized architectures. Along this line of ideas, many investigators have proposed cellular~\cite{Fiete2006, Schiess2016, Kusmierz2017, Sacramento2018, Richards2019, Kornfeld2020, Liu2020} and circuit level mechanisms~\cite{OReilly1996, Roelfsema2005, Lillicrap2016, Roelfsema2018} to assign credit appropriately. In this review, we would like to advance the notion that the specialized hardware arrangement also happens at the system level and propose that the thalamus and its interaction with basal ganglia (BG) and the cortex serve as a system-level solution for these three types of credit assignment.




\begin{figure}[h!]
    \centering
    \includegraphics[width = 0.55\textwidth]{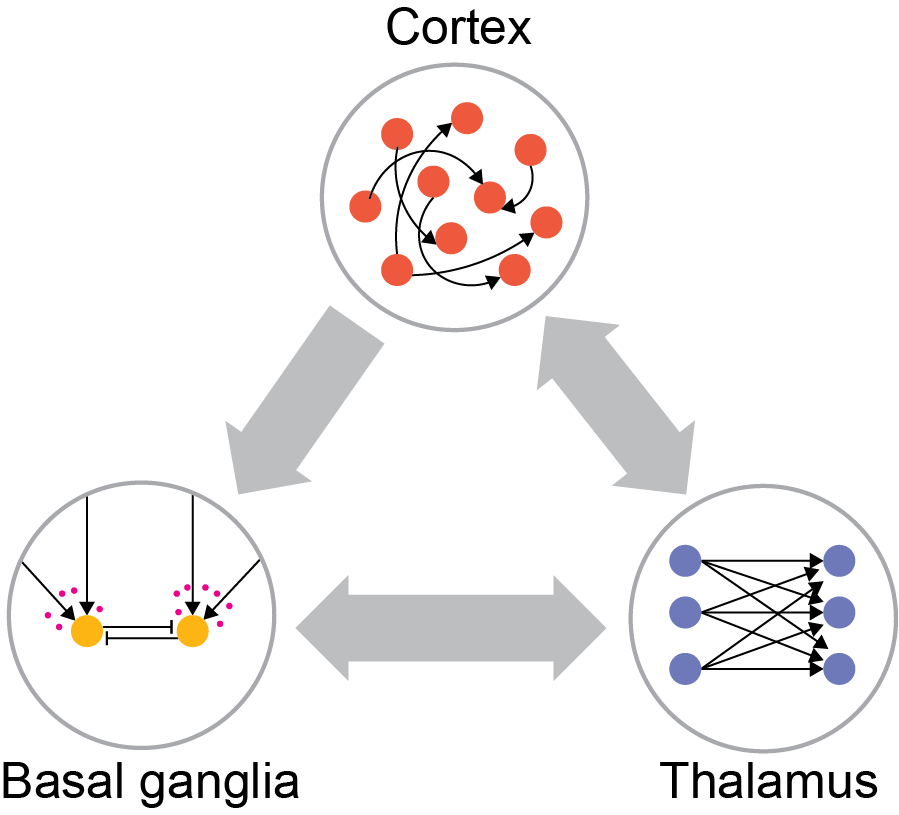}
    \caption{\textbf{Distinct architectures of cortex, thalamus and basal ganglia} Cortex is largely composed of excitatory neurons with extensive recurrent connectivity. Thalamus consists of mostly excitatory neurons without lateral connections. Basal ganglia consist of mostly inhibitory neurons driven by cortical and thalamic inputs, and the corticostriatal plasticity is modulated by dopamine.}
    \label{fig:system}
\end{figure}
\paragraph{A proposal: thalamocortical-basal ganglia interactions enable meta-learning to solve credit assignment.}

To motivate the notion of thalamocortical-basal ganglia interactions being a potential solution for credit assignment, we will start with a brief introduction. The cortex, thalamus and basal ganglia are the three major components of the mammalian forebrain -- the part of the brain to which high level cognitive capacities are attributed to~\cite{Alexander1986, Miller2000, Miller2001, Niv2009, Badre2010, Seo2012, Makino2016, Cox2019, Wolff2019}. Each of these components has its specialized internal architectures; the cortex is dominated by excitatory neurons with extensive lateral connectivity profiles~\cite{ Rakic2009, Fuster1997, Singer2019}, the thalamus is grossly divided into different nuclei harboring mostly excitatory neurons devoid of lateral connections~\cite{Jones1985, Sherman2005, Harris2019}, and the basal ganglia are a series of inhibitory structures driven by excitatory inputs from the cortex and thalamus~\cite{Gerfen2010, Nambu2011, Lanciego2012} (\autoref{fig:system}). A popular view within system neuroscience stipulates that BG and the cortex underwent different learning paradigms where BG is involved in reinforcement learning while the cortex is involved in unsupervised learning~\cite{Doya1999, Doya2020}. Specifically, the input structure of the basal ganglia known as the striatum is thought to be where reward gated plasticity takes place to implement reinforcement learning~\cite{Niv2009, Hikosaka2014, Cox2019, Bamford2018, Perrin2019, Kornfeld2020}. One such evidence is the high temporal precision of DA activity in the striatum. To accurately attribute the action that leads to positive RPE, DA is released into the relevant corticostriatal synapses. However, DA needs to disappear quickly to prevent the next stimulus-response combination from being reinforced. In the striatum, this elimination process is carried out by dopamine active transporter (DAT) to maintain a high temporal resolution of DA activity to support reinforcement learning~\cite{Garris1994, Ciliax1995, Cass1995}. In contrast, although the cortex also has dopaminergic innervation, cortical DAT expression is low and therefore DA levels may change at a timescale that is too slow to support reinforcement learning~\cite{Garris1994, Cass1995, Lapish2007, Seamans2010} but instead support other processes related to learning~\cite{Miller2001, Miller2001, Badre2010}. In fact, ample evidence indicates that cortical structures undergo Hebbian-like long term potentiation (LTP) and long term depression (LTD)~\cite{Kirkwood1996, Feldman2009, Cooke2010}. However, despite the unsupervised nature of these processes, cortical representations are task-relevant and include appropriate sensorimotor mappings that lead to rewards~\cite{Donahue2015, Tsutsui2016, Allen2017, Petersen2019, Enel2020, Jacobs2020}. How could this arise from an unsupervised process? One possible explanation is that basal ganglia activate the appropriate cortical neurons during behaviors and the cortical network collectively consolidates high reward sensorimotor mappings via Hebbian-like learning~\cite{Ashby2007, Andalman2009, Warren2011, Helie2015, Tesileanu2017}. Previous computational accounts of this process have emphasized a consolidation function for the cortex in this process, which naively would beg the question of why duplicate a process that seems to function well in the basal ganglia and perhaps include a lot of details of the associated experience?

\begin{figure}[h!]
    \centering
    \includegraphics[width = 0.9\textwidth]{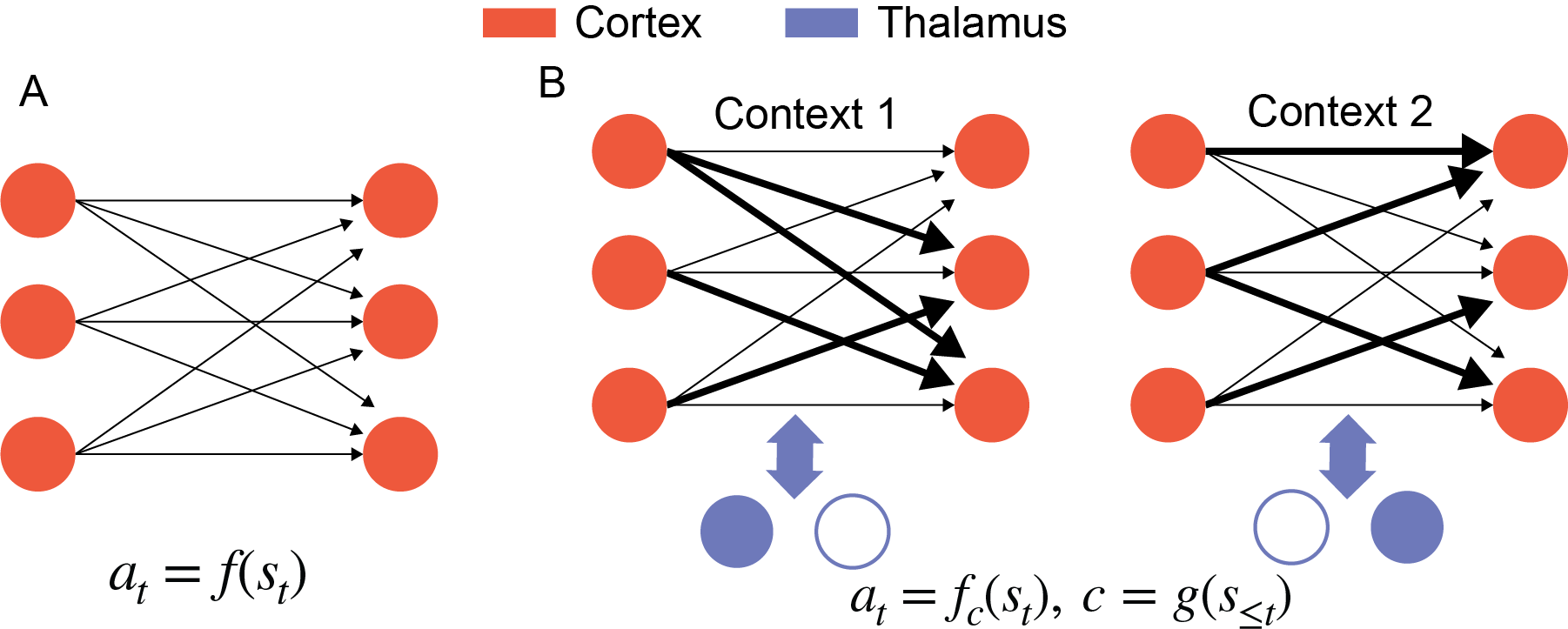}
    \caption{\textbf{Two views of learning in the cortex} A. One possible view is that the Hebbian cortical plasticity consolidates the sensorimotor mapping from BG to learn a stimulus-action mapping $a_t = f(s_t)$. B. We propose that thalamocortical systems do meta-learning by consolidating the teaching signals from BG to learn a context-dependent mapping $a_t = f_c(s_t)$ where the context $c$ is computed by past stimulus history and represented by different thalamic activities.}
    \label{fig:meta learning}
\end{figure}

The answer to this question is the core of our proposal. We propose that the learning process is not a duplication, but instead that the reinforcement process in the basal ganglia selects thalamic control functions that subsequently activate cortical associations to allow flexible mappings across different contexts  (\autoref{fig:meta learning}). 

To understand this proposition, we need to take a closer look at the involvement of these distinct network elements in task learning. Learning in basal ganglia happens in corticostriatal synapses where the basic form of reinforcement learning is implemented. Specifically, the coactivation of sensory and motor cortical inputs generates eligibility traces in corticostriatal synapses that get captured by the presence or absence of DA~\cite{Fiete2007, Fee2011, Kornfeld2020}. This RL algorithm is fast at acquiring simple associations but slow at generalization to other behaviors. On the other hand, the cortical plasticity operates in a much slower timescale but seems to allow flexible behaviors and fast generalization~\cite{Miller2000,Miller2001, Kim2011, Mante2013}. How does the cortex exhibit slow synaptic plasticity and flexible behaviors at the same time? An explanatory framework is meta-learning~\cite{Wang2018, Botvinick2019}, where the flexibility arises from network dynamics and the generalization emerges from slow synaptic plasticity across different contexts. In other words, synaptic plasticity stores a higher-order association between contexts and sensorimotor associations while the network dynamics switches between different sensorimotor associations based on this higher order association. However, properly arbitrating between synaptic plasticity and network dynamics to store such higher order association is a nontrivial task~\cite{Sohn2021}. We propose that the thalamocortical system learns these dynamics, where the thalamus provides control nodes that parametrize the cortical activity association space. Basal ganglia inputs to the thalamus learn to select between these different control nodes directly implementing the interface between weight adjustment and dynamical controls. Our proposal rests on the following three specific points. 

First, building on a line of the literature that shows diverse thalamocortical interaction in sensory, cognitive and motor cortex, we propose that thalamic output may be described as control functions over cortical computations. These control functions can be purely in the sensory domain like attentional filtering, in the cognitive domain like manipulating working memory or in the motor domain like preparation for movement~\cite{Tanaka2007, Saalmann2015, Wimmer2015, Zhou2016, Bolkan2017, Schmitt2017, Guo2017, Guo2017a, Rikhye2018, Mukherjee2020}. These functions directly relate thalamic activity patterns to different cortical dynamical regimes and thus offer a way to establish higher order association between context and sensorimotor mapping within the thalamocortical pathways. Second, based on previous studies on direct and indirect BG pathways that influence most cortical regions~\cite{Hunnicutt2016, Jiang2018, Nakajima2019, Peters2021}, we propose that BG hierarchically selects these thalamic control functions to influence activities of the cortex toward rewarding behavioral outcomes. Lastly, we propose that thalamocortical structure consolidate the selection of BG through a two timescales Hebbian learning process to enable meta-learning. Specifically, the faster corticothalamic plasticity learns the higher order association that enables flexible contextual switching with different thalamic patterns~\cite{Rikhye2018, Marton2018} while the slower cortical plasticity learns the shared representations that allow generalization to new behaviors. Below, we will go over the supporting literature that leads us to this proposal. 

\paragraph{More general roles of thalamocortical interaction and basal ganglia}
Classical literature has emphasized the role of the thalamus in transmitting sensory inputs to the cortex. This is because some of the better studied thalamic pathways are those connected to sensors on one end and primary cortical areas on another~\cite{HUBEL1961, Sherman1982, Reinagel1999, Usrey2000, Lien2018}. From that perspective, thalamic neurons being devoid of lateral connection transmit their inputs (e.g. from the retina in the case of the lateral geniculate nucleus (LGN)) to the primary sensory cortex (V1 in this same example case) and the input transformation (center-surround to oriented edges) occurs within the cortex~\cite{HUBEL1962, Hoffmann1972, Usrey2000, Lien2018}. In many cases, these formulations of thalamic ``relay" have generalized to how motor and cognitive thalamocortical interactions may be operating. However, in contrast to the classical relay view of the thalamus, more recent studies have shown diverse thalamic functions in sensory, cognitive and motor processing~\cite{Tanaka2007, Saalmann2015, Wimmer2015, Zhou2016, Bolkan2017, Schmitt2017, Guo2017, Guo2017a, Rikhye2018}. For example in mice, sensory thalamocortical transmission can be adjusted based on PFC-dependent, top-down biasing signals transmitted through non-classical basal ganglia pathways involving the thalamic reticular nucleus (TRN)~\cite{Wimmer2015, Phillips2016, Nakajima2019}. Interestingly, these task-relevant PFC signals themselves require long range interactions with the associative mediodorsal (MD) thalamus to be initiated, maintained and flexibly switched~\cite{Wimmer2015, Schmitt2017, Rikhye2018}. One can also observe nontrivial control functions in the motor thalamus. Motor preparatory activities in the anterior motor cortex (ALM) show persistent activities that predicted future actions. Interestingly, the motor thalamus also shows similar preparatory activities that predict future actions and by optogenetically manipulate the motor thalamus activities, the persistent activities in ALM quickly diminished~\cite{Guo2017}. Based on the above studies, we propose that the thalamus provides a set of control functions to the cortex. Specifically, cortical computations may be flexibly switched to different dynamical modes by activating a particular thalamic output that corresponds to that mode. 

On the other hand, the selective role of BG in motor and cognitive control also has dominated the literature because thalamocortical-basal ganglia interaction is the most well studied in frontal systems~\cite{Monchi2006,McNab2008,Seo2012,Makino2016, Cox2019}. However, classical and contemporary studies have recognized that all cortical areas, including primary sensory areas project to the striatum~\cite{Hunnicutt2016, Jiang2018, Peters2021}. Similarly, the basal ganglia can project to the more sensory parts of the thalamus through lesser-studied pathways to influence the sensory cortex~\cite{Hunnicutt2016,  Nakajima2019, Peters2021}. Specifically, a non-classical BG pathway projects to TRN which in turn modulates the activities of LGN to influence sensory thalamocortical transmission~\cite{Nakajima2019}. On the other hand, it has also been argued that BG are involved in gating working memory~\cite{McNab2008, Voytek2010}. This shows that BG has a much more general role than classical action and action strategy selection. Therefore, combining with our proposals on thalamic control functions, we propose that BG hierarchically selects different thalamic control functions to influence all cortical areas in different contexts through reinforcement learning.

Furthermore, there are series of the work that indicates the role of BG to guide plasticity in thalamocortical structures~\cite{Fiete2007,Andalman2009, Mehaffey2015, Helie2015, Tesileanu2017}. In particular, there is evidence that BG is critical for the initial learning and less involved in the automatic behaviors once the behaviors are learned across different species. In zebra finches, the lesion of BG in adult zebra finch has little effects on song production, but the lesion of BG in juvenile zebra finch prevents the bird from learning the song \cite{Sohrabji1990,Scharff1991, Fee2011}. Similar patterns can be observed in people with Parkinson's disease. Parkinson's patients who have a reduction of DA and striatal defects have troubles in solving procedural learning tasks but can produce automatic behaviors normally~\cite{Soliveri1997, Thomas1999, Asmus2008}. This behavioral evidence suggests that thalamocortical structures consolidate the learning from BG as the behaviors become more automatic. Furthermore, on the synaptic level, a songbird learning circuit also demonstrates this cortical consolidation motif~\cite{Mehaffey2015, Tesileanu2017}. In a zebra finch, the premotor nucleus HVC (a proper name) projects to the motor nucleus robust nucleus of the arcopallium (RA) to produce the song. On the other hand, RA also receives BG nucleus Area X mediated inputs from the lateral nucleus of the medial nidopallium (LMAN). The latter pathway is believed to be a locus of reinforcement learning in the songbird circuit. By burst stimulating both input pathways in different time lags, one can discover that HVC-RA and LMAN-RA underwent opposite plasticity~\cite{Mehaffey2015}. This suggests that the learning is gradually transferred from LMAN-RA to HVC-RA pathway~\cite{Fee2011,Mehaffey2015, Tesileanu2017}. This indicates a general role of BG as the trainer for cortical plasticity. We further propose that BG is the trainer in two different timescales for thalamocortical structures to enable meta-learning. The faster timescale trainer trains the corticothalamic connections to select the appropriate thalamic control functions in different contexts while the slower timescale trainer trains the cortical connections to form a task-relevant and generalizable representation.

\paragraph{The thalamocortical structure consolidates the BG selections on thalamic control functions in different timescales to enable meta-learning.}

\begin{figure}[h!]
    \centering
    \includegraphics[width= 0.9\textwidth]{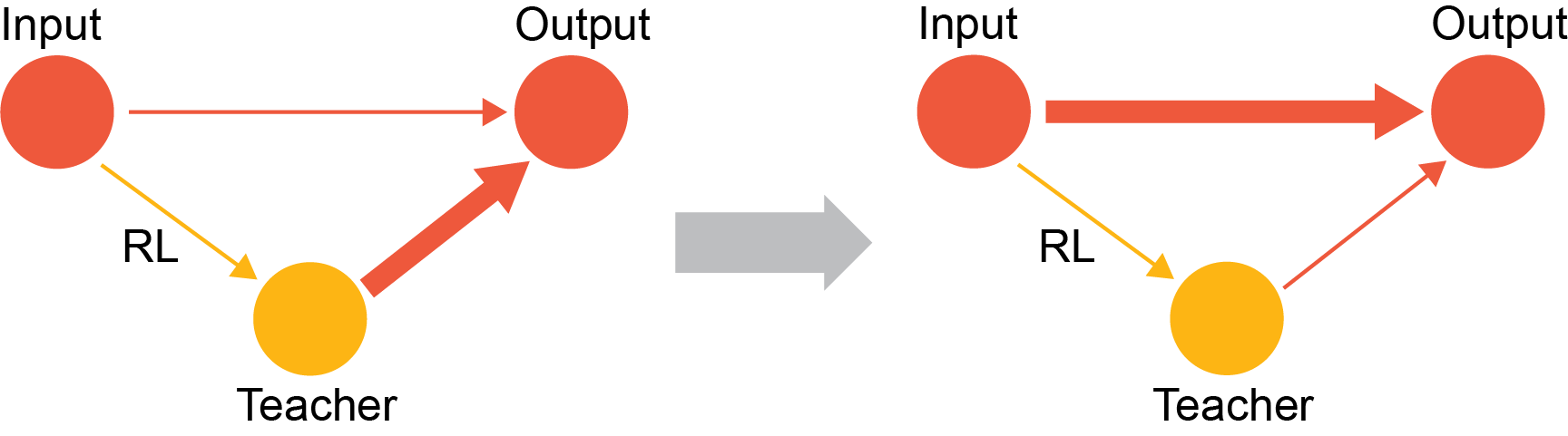}
    \caption{\textbf{A circuit motif for learning consolidation.} The outputs receive both student inputs and teacher inputs. Teacher inputs are learned from reinforcement learning. At the initial stage of learning, teacher inputs dominate the output. However, whenever the student inputs and the teacher inputs coincide, the student inputs are strengthened while the teacher inputs are weakened. Therefore, when the students learn the task, the learning is consolidated into the student connections and the teacher has little influence on the outputs.}
    \label{fig:motif}
\end{figure}

In this section, we will first propose a general network consolidation motif inspired by a songbird circuit mentioned above and then show how one can apply this motif in two different timescales in thalamocortical-basal ganglia interactions to enable meta-learning. We propose the following general network motif on learning consolidation (\autoref{fig:motif}). In this motif, there are three groups of neurons, input, output and teacher neurons. This is an abstract depiction that could map on multiple circuit scenarios (i.e. what the actual input, output and teacher circuits are). However, as a way of example, we can imagine an implementation of the input and output being intracortical and the teacher being the basal ganglia. As a general rule, the output receives both input and teacher connections, with the former undergoing unsupervised plasticity and the latter undergoing reinforcement learning. A key idea that we propose, is that unsupervised plasticity requires coactivation of both the direct and indirect pathways depicted in \autoref{fig:motif}. Specifically, the direct connections strengthen when it coincides with the teacher inputs (Hebbian-like) while the teacher connections weaken under the same conditions (anti-Hebbian-like). Therefore, at the initial stage of learning, the indirect pathway will have a strong influence on the output, which will lessen as the direct pathway is strengthened and thus the behaviors become less teacher dependent. We describe a basic mathematical setup of the network motif in~\autoref{box:motif}.

\begin{figure}[h!]
\begin{pabox}[label={box:motif},nameref={Box}]{A network motif for structural credit assignment with RL}
In this box, we will introduce a basic setup of this motif and the intuition on how it works. Given the activity of the input neurons $x(t)\in\Real^n$, the teacher neurons $y(t)\in \Real^m$, the output neurons $o(t)\in \Real^m$ and the input-teacher synapses $W\in \Real^{m\times n}$, input- output synapses $M\in \Real^{m\times n}$ and teacher-output synapses $w\in \Real^m$, we have the following relationship
\begin{equation}
o(t) = Mx(t) + w\cdot y(t),  
\end{equation}
\begin{equation}
g(t) = Wx(t),\ y_i(t) = g_i(t) \text{ if }i=a_t^*,\ y_i(t) = 0 \text{ otherwise}   
\end{equation}
where $\cdot$ is the entrywise product and $a_t^*$ is the action selected by the teacher. And the animals select the action $a_s^*$ as the behavior output based on the output activity. One possible action selection mechanism could be
\begin{equation}
\P(a_t^* = i) = \frac{e^{g_i(t)}}{\sum_{j=1}^me^{g_j(t)}},\ \P(a_s^* = i) = \frac{e^{s_i(t)}}{\sum_{j=1}^me^{s_j(t)}}.
\end{equation}
The plasticity rule of the input-teacher synapse can be described by reinforcement learning
\begin{equation}
\Delta W_{a_s^*} \propto (r_t - s_{a_s^*}(t))x(t),\ \Delta W_i = 0 \text{ if }i\neq a_s^*.
\end{equation}
The input-output synapses got strengthen if the input-output and teacher-output synapses coactivate the output. Below is one implementation of this idea
\begin{equation}
\Delta M_{a_t^*} \propto w_{a_t^*}t_{a_t^*}M_{a_t^*}\cdot x(t).
\end{equation} And furthermore, the teacher-output synapses undergo the opposite plasticity
\begin{equation}
\Delta w_{a_t^*} \propto - w_{a_t^*}t_{a_t^*} M_{a_t^*}^\top x(t),
\end{equation}
so the net contribution to the output is balanced. The decision of the output will rely on the teacher less as the input-output synapses learn the right sensorimotor association from the teacher. 
\end{pabox}
\end{figure}

\begin{figure}[h!]
    \centering
    \includegraphics[width=0.65\textwidth]{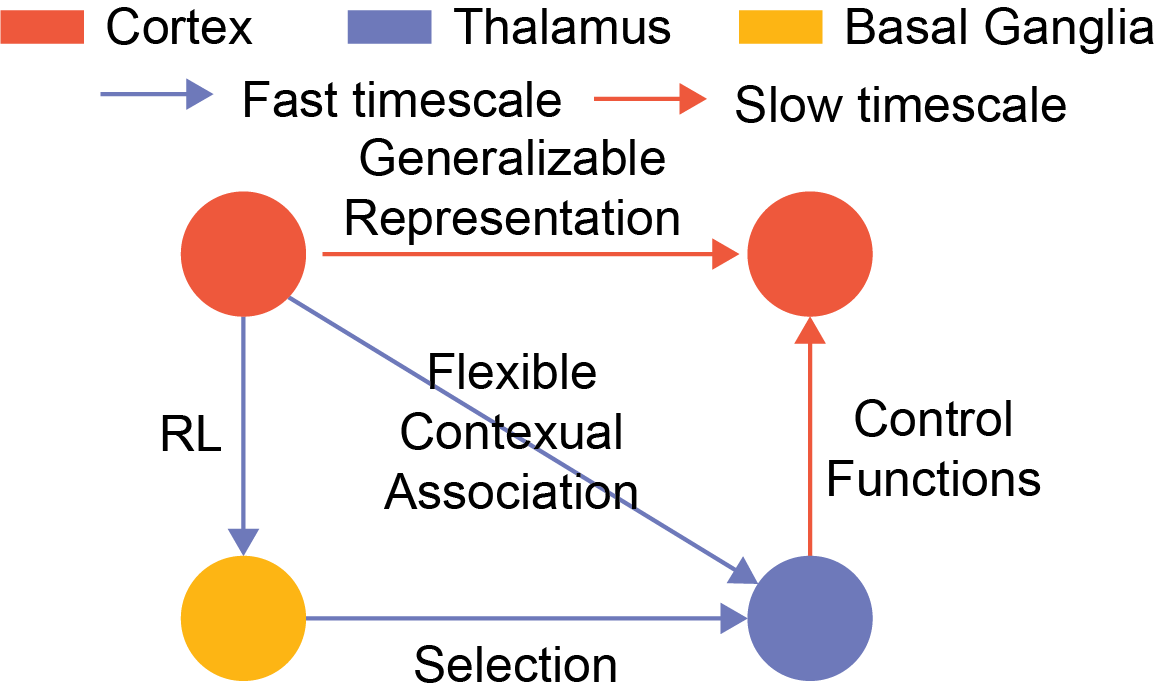}
    \caption{\textbf{Two timescales learning in thalamocortical structures} We propose that one can learn the thalamocortical structure to enable meta-learning by applying the general network motif in two different timescales. First, one can learn the corticothalamic connections by applying the motif on the blue loop with a faster timescale. This allows the network to consolidate flexible switching behaviors. Second, one can learn the cortical connections by applying the motif on the orange loop in a slower timescale. This allows cortical neurons to develop a task-relevant shared representation that can generalize across contexts.}
    \label{fig:thalamocortical loop}
\end{figure}

From the songbird example, we see how thalamocortical structures can consolidate simple associations learned through the basal ganglia. To enable meta-learning, we propose that this general direct/indirect connectivity idea operates over two different timescales within thalamocortical-basal ganglia interactions (\autoref{fig:thalamocortical loop}). First, combining the idea of thalamic outputs as control functions over cortical network activity patterns and the basal ganglia selecting such functions, we frame learning in basal ganglia as a process that connects contextual associations (higher order) with the appropriate dynamical control that maximizes reward at the sensorimotor level (lower order). Under this framing, corticothalamic plasticity consolidates the higher order association within a fast timescale. This allows flexible switching between different thalamic control functions in different contexts. On the other hand, the cortical plasticity consolidates the sensorimotor association over a slow timescale to allow shared representation that can generalize across different contexts. As the thalamocortical structures learn the higher order association, the behaviors become less BG-dependent and the network is able to switch between different thalamic control functions to induce different sensorimotor mappings in different contexts. By having two learning timescales, animals can conceivably both adapt quickly in changing environments with fast learning of corticothalamic connections while maintaining the important information across the environment in the cortical connections.

Some anatomical observations support this idea. Classically, thalamocortical neurons are classified into matrix and core~\cite{Jones2001,Sherman2005}. Matrix has a more modulatory role to the cortical dynamics in a diffusive projection while core has a driver role to the cortical dynamic in a topographically restricted dense projection. Most importantly, matrix has innervation to striatum while core does not. This indicates that matrix might serve as the role of control functions in the faster consolidation loop with the feedback to striatum to conduct credit assignment. On the other hand, core might be more involved in the slower consolidation loop with the feedback to striatum coming from the cortex to train the common cortical representation across contexts. 

In summary, this two timescales network consolidation scheme provides a general way for BG to guide plasticity in the thalamocortical architecture to enable meta-learning and thus solves structural credit assignment as a special case. Along these lines, experimental evidence supports the notion that when faced with multi-sensory inputs, the BG can selectively disinhibit a modality-specific subnetwork of the thalamic reticular nucleus (TRN) to filter out the sensory inputs that are not relevant to the behavior outcomes and thus solve the structural credit assignment problem. 

In the discussion above, we discuss our proposal under a general formulation of thalamic control functions. In the next sections, we will specify other thalamic control functions suggested by recent studies and observe how they can solve contextual and temporal credit assignments under this framework as well.

\paragraph{The Thalamus selectively amplifies functional cortical connectivity as a solution to contextual credit assignment and catastrophic forgetting.} One of the pitfalls of the artificial neural network is catastrophic forgetting. If one trains an artificial neural network on a sequence of tasks, the performance on the older task will quickly deteriorate as the network learns the new task \cite{McCloskey1989, French1999, Kumaran2016,Kemker2018,  Parisi2019}. On the other hand, the brain can achieve \textit{continual learning}, the ability to learn different tasks in different contexts without catastrophic forgetting, and thus solves contextual credit assignment~\cite{Lewkowicz2014, Murray2016,Power2017, Zenke2017}. There are three main approaches in machine learning in dealing with catastrophic forgetting. First, one can use the regularization method to mostly update the weights that are less important to the prior tasks \cite{Kirkpatrick2017, Zenke2017a, Fernando2017, Li2018, Jung2018, Maltoni2019}. This idea is inspired by experimental and theoretical studies on how synaptic information is selectively protected in the brain \cite{Fusi2005, Yang2009, Cichon2015, Hayashi2015, Benna2016}. However, it is unclear how to biologically compute the importance of each synapse to prior tasks nor how to do global regularization locally. Second, one can also use a dynamic architecture in which the network expands the architecture by allocating a subnetwork to train with the new information while preserving old information \cite{Xiao2014,Rusu2016, Draelos2017, Cortes2017}. However, this type of method is not scalable since the number of neurons needs to scale linearly with the number of the task. Lastly, one can use a memory buffer to replay past tasks to avoid catastrophic forgetting by interleaving the experience of the past tasks with the experience of the present task \cite{McClelland1995, Kumaran2016, Shin2017, Kemker2018a}. However, this type of method cannot be the sole solution as the memory buffer needs to scale linearly with the number of the tasks and potentially the number of the trials.

\begin{figure}[h!]
    \centering
    \includegraphics[width = 0.9\textwidth]{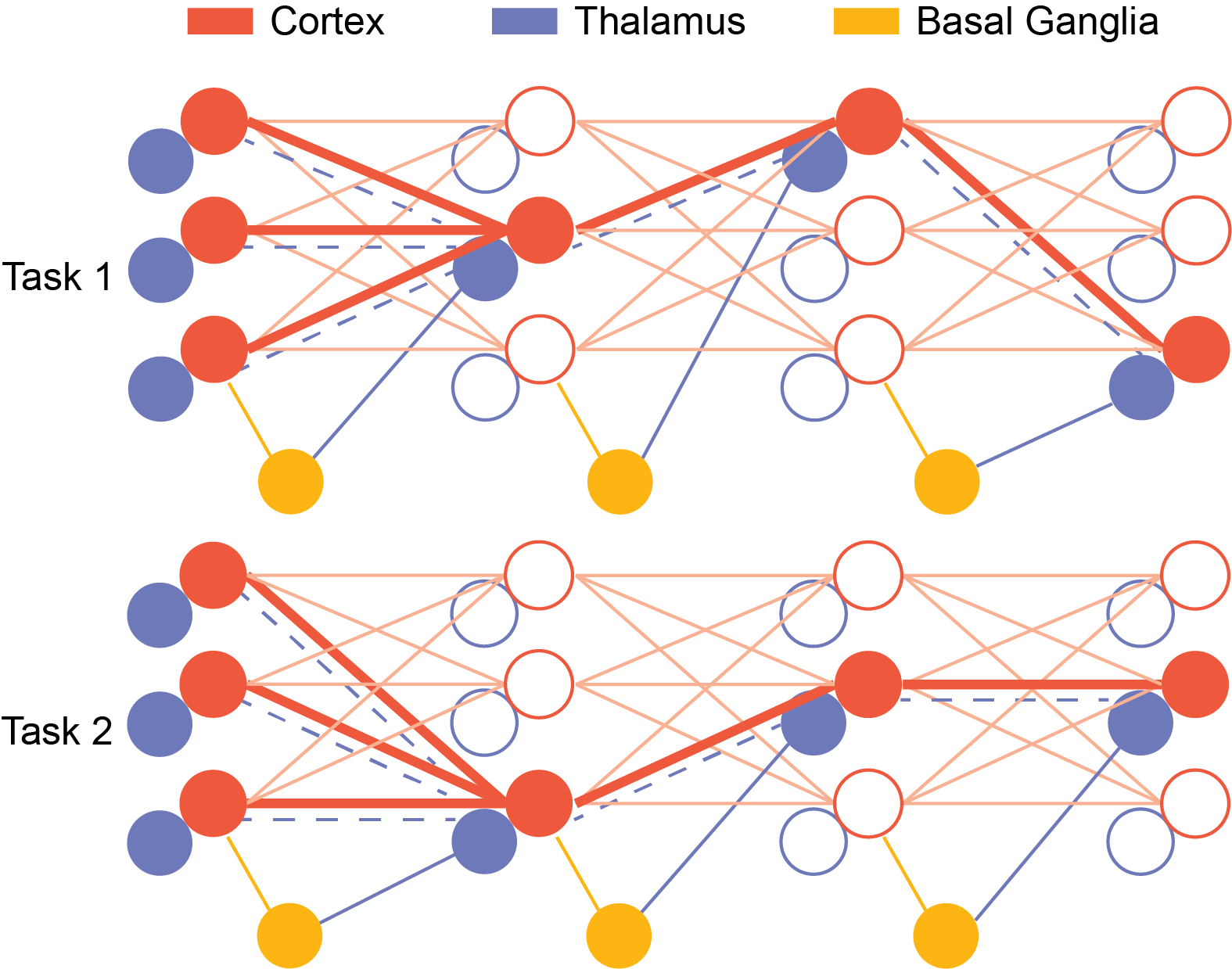}
    \caption{\textbf{A thalamocortical architecture with interaction with BG for contextual credit assignment} During task execution, BG selects thalamic neurons that amplify the relevant cortical subnetwork. This protects other parts of the network that are important for another context from being overwritten. When the other task comes, BG selects other thalamic neurons and since the synapses are protected from the last task, animals can freely switch from different tasks without forgetting the previous tasks. Furthermore, as the corticothalamic synapses learn how to select the right thalamic neurons in a different context (blue dash line), task execution can become less BG dependent.}
    \label{fig:continual learning}
\end{figure}

We propose that the thalamus provides another way to solve contextual credit assignment and catastrophic forgetting via selectively amplifying parts of the cortical connections in different contexts (\autoref{fig:continual learning}). Specifically, we propose that a population of thalamic neurons topographically amplify the connectivity of cortical subnetworks as their control functions. During a behavioral task, BG selects subsets of the thalamus which selectively amplify the connectivity of cortical subnetworks. Because of the reinforcement learning in BG, the subnetwork that is the most relevant to the current task will be more preferentially activated and updated. By selecting only the relevant subnetwork to activate in one context, the thalamus protects other subnetworks which can have useful information in another context from being overwritten. The corticothalamic structures can then consolidate these BG-guided flexible switching behaviors via our proposed network motif and the switching becomes less BG-dependent. Furthermore, our proposed solution has implications on generalization as well. Different tasks can have principles in common that can be transferred. For example, although the rules of chess and Go are very different, players in both games all need to predict what the other players are going to do and counterattack based on the prediction. Since BG selects the subnetwork at each hierarchy that is most relevant to the current tasks, in addition to selecting different subnetworks to prevent catastrophic forgetting, BG can also select subnetworks that are beneficial to both tasks as well to achieve generalization. Therefore, the cortex can develop a modular hierarchical representation of the world that can be easily generalized.

The idea of protecting relevant information from the past tasks to be overwritten has been applied before computationally and has decent success in combating catastrophic forgetting in deep learning \cite{Kirkpatrick2017}. Experimentally, we also have found thalamic neurons selectively amplify the cortical connectivity to solve the contextual credit assignment problem. In a task where the mice need to switch between different sets of task cues that guided the attention to the visual or auditory target, the performance of the mice does not deteriorate much after switching to the original context which is an indication of contextual credit assignment  \cite{Rikhye2018}. Electrophysiological recording of PFC and mediodorsal thalamic nucleus (MD) neurons, we discovered that PFC neurons preferentially code for the rule of the attention while MD neurons preferentially code for the contexts of different sets of the cues. Thalamic neurons that encode the task-relevant context translate this neural representation into the amplification of cortical activity patterns associated with that context (despite the fact that cortical neurons themselves only encode the context implicitly). These experimental observations are consistent with our proposed solution: by incorporating the thalamic population that can selectively amplify connectivity of cortical subnetworks, the thalamus and its interaction with cortex and BG solve the contextual credit assignment problem and prevent catastrophic forgetting.

\paragraph{Thalamus's role as a flexible gate for working memory makes temporal credit assignment more efficient.}Finally, in addition to structural credit assignment and contextual credit assignment, an animal also needs to solve temporal credit assignment, the ability to attribute the distant reward to the relevant past actions. One well-known mechanism to solve this problem is through temporal difference reward prediction error in the ventral tegmental area (VTA) of BG. Since the TD RPE captures the value function of discounted future reward, an organism does not need to wait until reward delivery to be able to credit the actions that lead to positive TD RPE. However, TD learning assumes that the state transition and the reward depend only on the current state and the current action while in a natural environment, many rewards depend on not only the current sensory inputs but also the past sensory inputs. One possible solution is to store all past sensory inputs and actions in the current state. However, this is highly inefficient since not all past information is relevant to the task. Therefore, the ability to store the relevant sensory inputs and actions in the working memory is important to solve temporal credit assignments in a natural environment. One such solution is LSTM network \cite{Hochreiter1997, Yu2019, VanHoudt2020} which selectively maintains past sensory inputs in the hidden memory. However, since one needs to use backpropagation to train the network, based on everything we have discussed so far, the brain ought to have an alternative solution.

\begin{figure}[h!]
    \centering
    \includegraphics[width=0.75\textwidth]{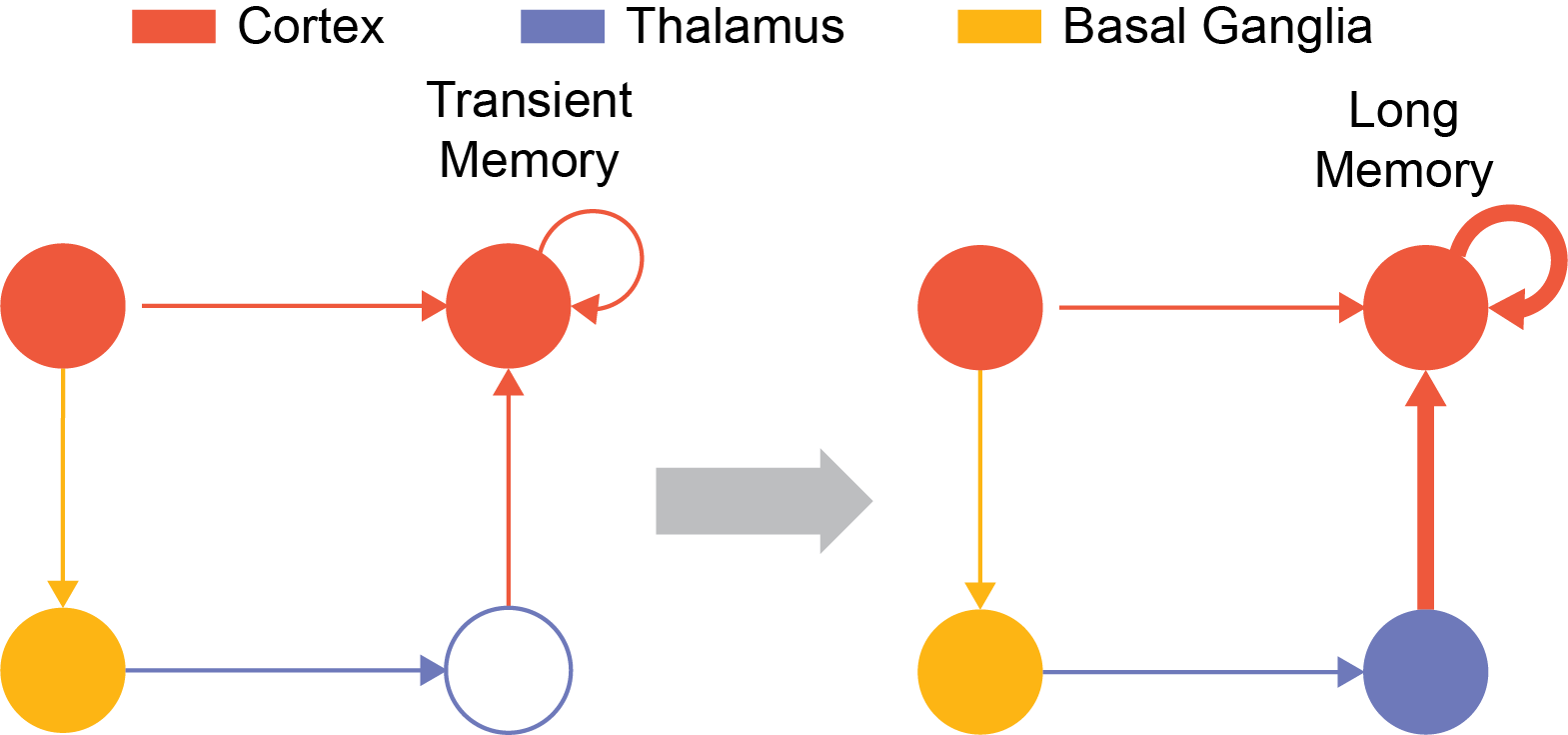}
    \caption{\textbf{BG uses reinforcement learning to adjust the time constant of the working memory} By inhibiting the specific thalamus that amplifies the connectivity of the working memory, BG gates the information from storing in the working memory. On the other hand, by disinhibiting the thalamus, BG selectively adjusts the time constant of the working memory.}
    \label{fig:working memory}
\end{figure}

We propose that the thalamus is also an integral component of the brain's solution to temporal credit assignment. By selectively changing the effective time constant of the working memory in cortical recurrent networks, thalamic outputs can precisely accomplish this computational objective. Specifically, when the sensory inputs arrive at prefrontal cortex (PFC), BG uses reinforcement learning to set thalamic activity patterns thereby selectively amplifying the appropriate cortical associations relevant to the task's working memory component (\autoref{fig:working memory}). Mathematically, we can consider the following simplified model of PFC
\begin{equation}
\tau\frac{dr_{PFC}}{dt} = -r_{PFC} + f(inp_{PFC}),\ inp_{PFC} = (1+r_{MD})w_{PFC}r_{PFC} + r_{s} 
\end{equation}
where $\tau$ is the time constant, $r_{PFC}$ is the PFC activity, $f(x) = (1+e^{-x})^{-1}$ is the sigmoid function, $r_{MD}$ is the input from MD to PFC, $w_{PFC}$ is the recurrent weight in PFC and $r_{s}$ is the input from sensory cortex. By linearizing the dynamic, we can derive the effective time constant of the working memory as
\begin{equation}
\tau_{eff} = \frac{\tau}{1 - (1+r_{MD})w_{PFC}f(inp_{PFC})(1-f(inp_{PFC}))}
\end{equation}
where larger $r_{MD}$ creates longer effective time constant in memory. Because of the reinforcement learning in BG to set the activity of $r_{MD}$, PFC is able to selectively store the task relevant sensory inputs into working memory. Furthermore, as the learning consolidates to corticothalamic connections, one then needs less BG dependency on adjusting the time constant of the working memory. By selectively maintaining the task-relevant information in working memory, one is able to credit the past information at the time when the reward is delivered.

The idea that the coordination between PFC, thalamus and BG selectively gates working memory has been explored computationally before and shows success in simple working memory tasks \cite{OReilly2006}. Experimentally, it is also well known that both thalamus~\cite{Bolkan2017,Guo2017, Schmitt2017} and basal ganglia~\cite{McNab2008, Voytek2010} are involved in cortical working memory. In our proposal, we further refine the idea by allowing thalamic control functions to fine-tune the time constant of the working memory through amplifying the cortical connectivity~\cite{Schmitt2017, Rikhye2018, Mukherjee2020}. By incorporating these thalamic populations that selectively adjust the effective time constant of cortical working memory into our framework, the thalamocortical-basal ganglia system can flexibly maintain the relevant sensory inputs to solve temporal credit assignment efficiently.  
 

\paragraph{Summary} In summary, in contrast to the traditional relay view of the thalamus,  we propose that thalamocortical interaction is the locus of meta-learning where the thalamus provides cortical control functions, such as sensory filtering, working memory gating or motor preparation, that parametrize the cortical activity association space. Furthermore, we propose a two timescale learning consolidation framework where BG hierarchically selects these thalamic control functions to enable meta-learning, solving the credit assignment problem. The faster plasticity learns contextual associations to enable rapid behavioral flexibility while the slower plasticity establishes cortical representation that generalizes. Additionally, by considering the ability of the thalamus to selectively amplify functional cortical connectivity, the thalamocortical-basal ganglia network is able to flexibly learn context-dependent association without catastrophic forgetting while generalizing to the new contexts. Lastly, by considering thalamic control functions in adjusting the time constant of cortical working memory, our framework provides an efficient solution to temporal credit assignment as well.


\bibliography{mybib}
\bibliographystyle{ieeetr}

\end{document}